\documentclass[11pt]{amsart}
\usepackage{amscd}
\usepackage{amsmath}
\usepackage{graphicx}
\usepackage{amsfonts}
\usepackage{amssymb}
\begin{document}

\title[Fidelity of states in infinite dimensional quantum systems]
{Fidelity of states in infinite dimensional quantum systems}

\author{Jinchuan Hou}
\address{Department of
Mathematics\\
Taiyuan University of Technology\\
 Taiyuan 030024,
  P. R. China}
\email{jinchuanhou@yahoo.com.cn, houjinchuan@tyut.edu.cn}
\author{Xiaofei Qi}
\address
{Department of Mathematics\\
 Shanxi University\\
  Taiyuan 030006\\
   P. R. China} \email{qixf1980@126.com}


\thanks{This work is partially supported by  Research Fund for the Doctoral Program of Higher Education of China (20101402110012),
 Tianyuan Funds of China (11026161) and
Foundation of Shanxi University.}

\begin{abstract}

In this paper we discuss the fidelity of states in infinite
dimensional systems,  give an elementary proof of the infinite
dimensional version of Uhlmann's theorem, and then, apply it to
generalize several properties of the fidelity from finite
dimensional case to infinite dimensional case. Some of them are
somewhat different from those for finite dimensional case.

\end{abstract}
\maketitle

\section{Introduction}

In quantum mechanics, a quantum system is
 associated with a separable complex Hilbert space $H$, i.e.,
the state space.  A quantum state is described as a density operator
$\rho\in{\mathcal T}(H)\subseteq{\mathcal B}(H)$ which is positive
and has trace 1, where ${\mathcal B}(H)$ and ${\mathcal T}(H)$
denote the von Neumann algebras of all bounded linear operators and
the trace-class of all operators $T$ with $\|T\|_{\rm Tr}={\rm
Tr}((T^\dagger T)^{\frac{1}{2}})<\infty$, respectively.  $\rho$ is a
pure state if $\rho^2=\rho$; $\rho$ is a mixed state if
$\rho^2\not=\rho$. Let us denote by ${\mathcal S}(H)$ the set of all
states acting on $H$.

Recall also that the fidelity of states $\rho$ and $\sigma$ in
${\mathcal S}(H)$ is defined to be
$$ F(\rho,\sigma )={\rm Tr} \sqrt{\rho^{1/2}\sigma\rho^{1/2}}.\eqno(1.1)$$
Fidelity is a very useful measure of closeness between two states
and has several nice properties including the Uhlmann's theorem.

Uhlmann and co-workers developed Eq.(1.1) by the transition
probability in the more general context of the representation theory
of C*-algebras \cite{U,A,AU}. The result in \cite{U} (also ref.
\cite{U2}) implies that, if $\dim H<\infty$, then the equality
$$F(\rho,\sigma )=\max |\langle\psi|\phi\rangle|, \eqno(1.2)$$ holds, where the
maximization is over all purifications $|\psi\rangle$ of $\rho$ and
$|\phi\rangle$ of $\sigma$ into a larger system of $H\otimes H$.
This result is then referred as the  Uhlmann's theorem. Eq.(1.2)
does not provide a calculation tool for evaluating the fidelity, as
does Eq.(1.1). However, in many instances, the properties of the
fidelity are more easily deduced using Eq.(1.2) than Eq.(1.1). For
example, Eq.(1.2) makes it clear that $0\leq
F(\rho,\sigma)=F(\sigma,\rho)\leq 1$; $F(\rho,\sigma)=1$ if and only
if $\rho=\sigma$.

In \cite{J}, Jozsa presented an elementary proof of the Uhlmann's
theorem without involving the representation theory of  C*-algebras.
In this paper we will consider the fidelity of states in infinite
dimensional systems,  give an elementary proof of the infinite
dimensional version of Uhlmann's theorem, and then, apply it to
generalize several properties of the fidelity from finite
dimensional case to infinite dimensional case. Of course, not all
results for finite dimensional case can be generalized fully to
infinite dimensional case. For example, in the finite dimensional
case, it is known that $F(\rho,\sigma)=\min_{\{E_m\}} F(p_m,q_m),$
where the minimum is over all POVMs (positive operator-valued
measure) $\{E_m\}$, and $p_m={\rm Tr}(\rho E_m)$, $q_m={\rm
Tr}(\sigma E_m)$ are the probability distributions for $\rho$ and
$\sigma$ corresponding to the POVM $\{E_m\}$. However, this is not
true for infinite dimensional case. What we have is that
$F(\rho,\sigma)=\inf_{\{E_m\}} F(p_m,q_m)$. The infimum attains the
minimum if and only if $\rho$ and $\sigma$ meet certain condition.

Let $H$ be a complex Hilbert space, $A\in{\mathcal B}(H)$ and
$T\in{\mathcal T}(H)$. It is well known from the operator theory
that $|{\rm Tr}(AT)|\leq  \|AT\|_{\rm Tr}\leq \|A\|\|T\|_{\rm Tr}$.
This fact  will be used frequently in this paper.

\section{Infinite dimensional version of the Uhlmann's theorem and an elementary proof}

Recall that an operator $V\in{\mathcal B}(H)$ is called an isometry
if $V^\dag V=I$; is called a co-isometry if $VV^\dag =I$. If $\dim
H=\infty$ and $T\in{\mathcal B}(H)$, then, by the polar
decomposition, there exists an isometry or a co-isometry $V$ such
that $T=V|T|$, where $|T|=(T^\dag T)^{1/2}$. Generally speaking, $V$
may not be unitary. In fact, there exists a unitary operator $U$
such that $T=U|T|$ if and only if $\dim \ker T=\dim\ker T^\dag$.
However, the following lemma says that it is the case if $T$ is a
product of two positive operators.

{\bf Lemma 2.1.} {\it Let $H$ be a Hilbert space and
$A,B\in{\mathcal B}(H)$. If $A\geq 0$ and $B\geq 0$, then there
exists a unitary operator $V\in{\mathcal B}(H)$ such that
$AB=V|AB|$.}

{\bf Proof.} We need only to show that $\dim\ker AB=\dim\ker BA$ if
both $A$ and $B$ are positive operators.

Note that, since $A\geq 0$ and $B\geq 0$, we have
$$\ker AB=\ker B\oplus \ker A\cap (\ker B)^\bot \eqno(2.1)$$
and
$$\ker BA=\ker A\oplus \ker B\cap (\ker A)^\bot .\eqno(2.2)$$

Obviously, if $\dim \ker A=\dim\ker B=\infty$, then $\dim\ker
AB=\dim\ker BA=\infty$; if $A$ (or $B$) is injective, then $\dim\ker
AB=\dim\ker BA=\dim\ker B$ (or $\dim\ker AB=\dim\ker BA=\dim\ker
A$).

Assume that $\dim\ker A<\infty$ and $\dim\ker B=\infty$. By
Eqs.(2.1)-(2.2) we need only to check that $\dim \ker B\cap (\ker
A)^\bot =\infty$. This is equivalent to show the following
assertion.

{\bf Assertion.}  If $B\geq 0$ and $\dim\ker B=\infty$, then, for
any subspace $M\subset H$ with $\dim M^\bot< \infty$, $\dim\ker
P_MBP_M|_M=\infty$.

In fact, by the space decomposition $H=M\oplus M^\bot$, we may write
$B=\left(\begin{array}{cc}B_{11}
&B_{12}\\B_{12}^\dag&B_{22}\end{array}\right),$ where
$B_{11}=P_MBP_M|_M$. Since $B\geq 0$, there exists some contractive
operator $D$ such that $B_{12}=B_{11}^{1/2}DB_{22}^{1/2}$ (for
example, see \cite{H1}). Thus
$$\ker B=\ker B_{11}\oplus \ker B_{22}\oplus L,$$
where $$L= \{|x\rangle\oplus |y\rangle : |x\rangle\in (\ker
B_{11})^\bot, \ |y\rangle\in(\ker B_{22})^\bot,\
B_{11}|x\rangle+B_{12}|y\rangle=0\ \mbox {and}\ B_{12}^\dag
|x\rangle+B_{22}|y\rangle=0\}.$$ Note that $\dim\ker B_{22}<\infty$
and $\dim L\leq \dim (\ker B_{22})^\bot<\infty$, we must have
$\dim\ker B_{11}=\infty$.

Finally, assume that both $\ker A$ and $\ker B$ are finite
dimensional. With respect to the space decomposition $H=(\ker
A)^\bot\oplus\ker A$, we have
$$A=\left(\begin{array}{cc} A_1 &0\\0&0\end{array}\right) \quad \mbox{and}\quad
B=\left(\begin{array}{cc}B_{11}
&B_{12}\\B_{12}^\dag&B_{22}\end{array}\right).$$ As $A_1$ is
injective with dense range,
$$AB=\left(\begin{array}{cc} A_1B_{11} &A_1B_{12}\\0&0\end{array}\right) \quad \mbox{and}\quad
BA=\left(\begin{array}{cc}B_{11}A_1 &0\\B_{12}^\dag
A_1&0\end{array}\right),$$ we see that
$$\begin{array}{rl}\ker AB=&\{|x\rangle\oplus |y\rangle : |x\rangle\in (\ker A)^\bot,\ |y\rangle\in\ker A,\
B_{11}|x\rangle+B_{12}|y\rangle=0\}\\
=&(\ker B_{11}\oplus\ker B_{12})\\ &+\{|x\rangle\oplus |y\rangle:
|x\rangle\in(\ker B_{11})^\bot,\ |y\rangle\in(\ker B_{12})^\bot,\
B_{11}|x\rangle+B_{12}|y\rangle=0\}\end{array}
$$ and
$$\ker BA=\ker A \oplus \{|x\rangle : |x\rangle\in (\ker A)^\bot\cap\ker
B_{11}\}=\ker A\oplus\ker B_{11}.$$ Since $\dim \{|x\rangle\oplus
|y\rangle: |x\rangle\in(\ker B_{11})^\bot,\ |y\rangle\in(\ker
B_{12})^\bot,\ B_{11}|x\rangle+B_{12}|y\rangle=0\}\leq\dim(\ker
B_{12})^\bot$ and $\dim\ker B_{12}+\dim (\ker B_{12})^\bot=\dim\ker
A$, one gets
$$\dim \ker AB\leq \dim \ker BA.$$
Symmetrically, we have $\dim\ker BA\leq \dim\ker AB$, and therefore,
$\dim\ker AB=\dim\ker BA$. Complete the proof of the lemma.
\hfill$\Box$

If $\dim H<\infty$, then, for any $T\in{\mathcal B}(H)$, we have
$\|T\|_{\rm Tr}={\rm Tr}(|T|)=\max\limits_{U}\{{\rm Tr}(AU)\}$,
where the maximum is over all unitary operators. This result is not
valid even for trace-class operators if $\dim H=\infty$. The next
lemma says that the above result is true if the operator is a
product of two positive operators.

{\bf Lemma 2.2.} {\it Let $H$ be a complex Hilbert space and
$A,B\in{\mathcal B}(H)$. If $A,B$ are positive and $AB\in{\mathcal
T}(H)$, then
$$\|AB\|_{\rm Tr}={\rm Tr}(|AB|)=\max \{{\rm Tr}(ABU) :
U\in{\mathcal U}(H)\}, \eqno(2.3)$$ where ${\mathcal U}(H)$ is the
unitary group of all unitary operators in ${\mathcal B}(H)$.}

{\bf Proof.}  For any unitary operator $U\in{\mathcal U}(H)$, we
have
$$ |{\rm Tr}(ABU)|\leq\|U\|\|AB\|_{\rm Tr}=\|AB\|_{\rm Tr}={\rm
Tr}(|AB|).$$ On the other hand, by Lemma 2.1, there exists a unitary
operator $V$ such that $AB=V|AB|$. Thus $|AB|=V^\dag AB$ and
$$\|AB\|_{\rm Tr}={\rm
Tr}(|AB|)={\rm Tr}(V^\dag AB)={\rm Tr}(ABV^\dag).$$ Hence Eq.(2.3)
holds. \hfill$\Box$

{\bf Lemma 2.3.} {\it Let $H$, $K$ be separable infinite dimensional
complex Hilbert spaces and $A\in {\mathcal B}(H)$, $B\in {\mathcal
B}(K)$. Let $\{|i\rangle\}_{i=1}^\infty$,
$\{|i'\rangle\}_{i=1}^\infty$ be any  orthonormal bases  of $H$, $K$
respectively, and $U$ be the unitary operator defined by
$U|i\rangle=|i'\rangle$. For each positive integer $N$, let
$|m_N\rangle=\sum_{i=1}^N |i\rangle |i'\rangle$. If $A$ or $B$ is  a
trace-class operator, then, $$ \lim_{N\rightarrow\infty}\langle
m_N|A\otimes B|m_N\rangle ={\rm Tr}(UA^\dag U^\dag B).$$}

{\bf Proof.} Clearly, $UA^\dag U^\dag B\in{\mathcal T}(K)$ and
$${\rm Tr}(UA^\dag U^\dag B)=\sum _{i,j} \langle i'|UA^\dag U^\dag |j'\rangle\langle j'|B
|i'\rangle = \sum _{i,j} \langle i|A^\dag |j\rangle\langle j'|B
|i'\rangle, $$ which is absolutely convergent. Hence
$$ \lim _{N\rightarrow\infty}  \sum _{i,j=1}^N \langle i|A^\dag |j\rangle\langle j'|B
|i'\rangle={\rm Tr}(UA^\dag U^\dag B). \eqno(2.4)$$

On the other hand,
$$\langle m_N|A\otimes B|m_N\rangle =\sum_{i,j=1}^N \langle j|\langle j'|A\otimes B|i\rangle
|i'\rangle =\sum_{i,j=1}^N \langle j|A|i\rangle\langle
j'|B|i'\rangle = \sum _{i,j=1}^N \langle i|A^\dag |j\rangle\langle
j'|B |i'\rangle.
$$

So, by Eq.(2.4), one obtains that
$$
\lim_{N\rightarrow\infty}\langle m_N|A\otimes B|m_N\rangle ={\rm
Tr}(UA^\dag U^\dag B),$$ as desired. \hfill$\Box$

The following is the infinite dimensional version of the Uhlmann's
theorem. Recall that a unit vector $|\psi\rangle\in H\otimes K$ is
said to be a purification of a state $\rho$ on $H$ if $\rho={\rm
Tr}_K(|\psi\rangle\langle\psi|)$.

{\bf Theorem 2.4.}  {\it Let $H$ and $K$ be separable infinite
dimensional complex Hilbert spaces. For any states $\rho$ and
$\sigma$ on $H$, we have
$$ F(\rho,\sigma)=\max\{ |\langle \psi |\phi\rangle | :
|\psi\rangle\in{\mathcal P}_\rho,\ |\phi\rangle\in{\mathcal
P}_\sigma\}, $$ where ${\mathcal P}_\rho= \{|\psi\rangle\in H\otimes
K : |\psi\rangle \ \mbox{is a purification of}\ \rho\}$. }

{\bf Proof.} Assume that $\rho, \sigma\in{\mathcal S}(H)$. Then
there exist orthonormal bases of $H$, $\{|i_H\rangle\}_{i=1}^\infty$
and $\{|i'_H\rangle\}_{i=1}^\infty$ such that
$\rho=\sum_{i=1}^\infty p_i|i_H\rangle\langle i_H|$ and
$\sigma=\sum_{i=1}^\infty q_i|i'_H\rangle\langle i'_H|$ with
$\sum_{i=1}^\infty p_i=\sum_{i=1}^\infty q_i=1$. If $|\psi\rangle,
|\phi\rangle \in H\otimes K$ are purifications of $\rho$, $\sigma$,
respectively, then there exist orthonormal sets
$\{|i_K\rangle\}_{i=1}^\infty$ and $\{|i'_K\rangle\}_{i=1}^\infty$
in $K$ such that $|\psi\rangle =\sum_{i=1}^\infty
\sqrt{p_i}|i_H\rangle |i_K\rangle$ and $|\phi\rangle
=\sum_{i=1}^\infty \sqrt{q_i}|i'_H\rangle |i'_K\rangle$.

Pick any orthonormal bases $\{|i''_H\rangle\}_{i=1}^\infty$ of $H$
and $\{|i''_K\rangle\}_{i=1}^\infty$ of $K$. Let $U_H, U_K, V_H,
V_K$ be partial isometries defined by respectively
$$U_H|i''_H\rangle
=|i_H\rangle,\ U_K|i''_K\rangle =|i_K\rangle, \ V_H|i''_H\rangle
=|i'_H\rangle, \ V_K|i''_K\rangle =|i'_K\rangle \eqno(2.5)$$ for
each $i=1,2,\ldots $. For any integer $N>0$, let
$$|m_N\rangle
=\sum_{i=1}^N|i''_H\rangle|i''_K\rangle .$$ Then
$$|\psi_N\rangle=\sum_{i=1}^N
\sqrt{p_i}|i_H\rangle |i_K\rangle=\sum_{i=1}^N
\sqrt{\rho}(U_H\otimes U_K)|i''_H\rangle
|i''_K\rangle=(\sqrt{\rho}U_H\otimes U_K)|m_N\rangle$$ and
$$|\phi_N\rangle=\sum_{i=1}^N
\sqrt{q_i}|i'_H\rangle |i'_K\rangle=\sum_{i=1}^N
\sqrt{\sigma}(V_H\otimes V_K)|i''_H\rangle
|i''_K\rangle=(\sqrt{\sigma}V_H\otimes V_K)|m_N\rangle.$$ It follows
from Lemma 2.3 that
$$\begin{array}{rl}
|\langle \psi |\phi\rangle|= & \lim_{N\rightarrow\infty} |\langle
\psi_N |\phi_N\rangle|=\lim_{N\rightarrow\infty} |\langle
m_N|U_H^\dag\sqrt{\rho}\sqrt{\sigma}V_H\otimes U_K^\dag
V_K|m_N\rangle| \\= &|{\rm
Tr}(UV_H^\dag\sqrt{\sigma}\sqrt{\rho}U_HU^\dag U_K^\dag V_K)|\leq
\|U_HU^\dag U_K^\dag V_KUV_H^\dag\| {\rm
Tr}(|\sqrt{\sigma}\sqrt{\rho}|)
\\
\leq & {\rm Tr}(|\sqrt{\sigma}\sqrt{\rho}|)={\rm
Tr}\sqrt{\rho^{1/2}\sigma\rho^{1/2}}=F(\rho,\sigma),
\end{array} \eqno(2.6)
$$
where $U$ is the unitary operator defined by $U|i''_H\rangle
=|i''_K\rangle$. Therefore, we have proved that
$$ F(\rho,\sigma)\geq\sup\{ |\langle \psi |\phi\rangle | :
|\psi\rangle\in{\mathcal P}_\rho,\ |\phi\rangle\in{\mathcal
P}_\sigma\}.$$

Now, to complete the proof, it suffices to find
$|\psi\rangle\in{\mathcal P}_\rho$ and $|\phi\rangle\in{\mathcal
P}_\sigma$ such that $|\langle\psi|\phi\rangle|=F(\rho,\sigma)$.

By applying Lemma 2.1, we see that $\sqrt{\sigma}\sqrt{\rho}$ has a
polar decomposition
$\sqrt{\sigma}\sqrt{\rho}=U_0|\sqrt{\sigma}\sqrt{\rho}|$ with $U_0$
a unitary operator.

Let $\{|i_K\rangle\}_{i=1}^\infty$ be an orthonormal basis of $K$
and let $|\psi\rangle =\sum_{i=1}^\infty \sqrt{p_i}|i_H\rangle
|i_K\rangle$ and $|\phi\rangle =\sum_{i=1}^\infty
\sqrt{q_i}|i'_H\rangle |i_K\rangle$. Then $|\psi\rangle\in{\mathcal
P}_\rho$ and $|\phi\rangle\in{\mathcal P}_\sigma$. Let
$|i''_H\rangle=|i_H\rangle$, $|i''_K\rangle=|i_K\rangle$,
$i=1,2\ldots$. Then, by Eq.(2.5), $U_H=I$, $U_K=I$, $V_H$ is a
unitary operator determined by $V_H|i_H\rangle=|i'_H\rangle$.  Take
$|i'_K\rangle$ so that $V_K=UU_0^\dag V_HU^\dag$.  Then for such
choice of $|\psi\rangle$ and $|\phi\rangle$ we have

$$\begin{array}{rl}
|\langle \psi |\phi\rangle|= & \lim_{N\rightarrow\infty} |\langle
\psi_N |\phi_N\rangle|=\lim_{N\rightarrow\infty} |\langle
m_N|\sqrt{\rho}\sqrt{\sigma}V_H\otimes V_K|m_N\rangle| \\= &|{\rm
Tr}(UV_H^\dag\sqrt{\sigma}\sqrt{\rho}U^\dag V_K)|=| {\rm Tr}(U^\dag
V_KUV_H^\dag U_0|\sqrt{\sigma}\sqrt{\rho}|)|
\\
=&| {\rm Tr}(|\sqrt{\sigma}\sqrt{\rho}|)|=F(\rho ,\sigma ),
\end{array}
$$
completing the proof. \hfill$\Box$

By checking the proof of Theorem 2.4, it is easily seen that the
following holds.

{\bf Corollary 2.5.} {\it Let $H$ and $K$ be separable infinite
dimensional complex Hilbert spaces. For any states $\rho$ and
$\sigma$ on $H$, we have
$$ F(\rho,\sigma)=\max\{ |\langle \psi_0 |\phi\rangle | :
 |\phi\rangle\in{\mathcal
P}_\sigma\}=\max\{ |\langle \psi |\phi_0\rangle | :
 |\psi\rangle\in{\mathcal
P}_\rho\}, $$ where $|\psi_0\rangle$ is any fixed purification of
$\rho$ of the form $|\psi_0\rangle =\sum_{i=1}^\infty
\sqrt{p_i}|i_H\rangle |i_K\rangle$ with $\{ |i_K\rangle\}$ an
orthonormal basis of $K$ and $|\phi_0\rangle$ is any fixed
purification of $\sigma$ of the form $|\phi_0\rangle
=\sum_{i=1}^\infty \sqrt{q_i}|i_H'\rangle |i_K'\rangle$ with $\{
|i_K'\rangle\}$ an orthonormal basis of $K$.}

The fidelity is not a distance because it does not meet the
triangular inequality. However, like to the finite dimensional case,
by use of Theorem 2.4 and Corollary 2.5, one can show that the
are-cosine of fidelity is a distance.

{\bf Corollary 2.6.} {\it $A(\rho,\sigma)=:\arccos F(\rho,\sigma)$
is a distance on ${\mathcal S}(H)$.}

Several remarkable properties of fidelity in finite dimensional case
are still valid for infinite dimensional case. For instance,

{\bf Monotonicity of the fidelity} For any quantum channel
${\mathcal E}$, we have
$$ F({\mathcal E}(\rho), {\mathcal E}(\sigma))\geq F(\rho,\sigma).
\eqno(2.7)$$ Recall that a quantum channel is a
 completely positive and trace preserving linear map from ${\mathcal
 T}(H)$ into ${\mathcal T}(K)$.

{\bf Strong concavity of the fidelity} Let $p_i$ and $q_i$ be
probability distributions over the same index set, and $\rho_i$ and
$\sigma_i$ states also indexed by the same index set. Then
$$F(\sum_ip_i\rho_i, \sum_iq_i\sigma_i)\geq \sum_i
\sqrt{p_iq_i}F(\rho_i,\sigma_i). \eqno(2.8)$$

\section{Connection to the classical fidelity and trace distance}

If $\dim H<\infty$, the quantum fidelity is related to the classical
fidelity  by considering the probability distributions induced by a
measurement. In fact \cite[pp. 412]{NC}
$$F(\rho,\sigma)=\min\limits_{\{E_m\}} F(p_m,q_m), \eqno(3.1)$$
where the minimum is over all POVMs (positive operator-valued
measure) $\{E_m\}$, and $p_m={\rm Tr}(\rho E_m)$, $q_m={\rm
Tr}(\sigma E_m)$ are the probability distributions for $\rho$ and
$\sigma$ corresponding to the POVM $\{E_m\}$.

It is natural to ask whether or not  Eq.(3.1) is true if $\dim
H=\infty$? The following result is our answer.

For a positive operator $A\in{\mathcal B}(H)$, with respect to the
space decomposition $H=(\ker A)^\bot \oplus \ker A$,
$A=\left(\begin{array}{cc} A_1 &0\\ 0& 0\end{array}\right)$, where
$A_1:(\ker A)^\bot \rightarrow(\ker A)^\bot $ is injective and hence
$A_1^{-1}$ makes sense. In this paper, we always denote $A^{[-1]}$
for the may unbounded densely defined positive operator defined by
$A^{[-1]}=\left(\begin{array}{cc} A_1^{-1} &0\\ 0&
0\end{array}\right)$ with domain ${\mathcal D}(A^{[-1]})={\rm ran}
(A)\oplus \ker A$. Here ran($A)$ stands for the range of $A$.

{\bf Theorem 3.1.} {\it  Let $H$ be a separable infinite dimensional
complex Hilbert space. Then, for any states $\rho,\sigma\in
{\mathcal S}(H)$, we have
$$F(\rho,\sigma)=\inf\limits_{\{E_m\}} F(p_m,q_m), \eqno(3.2)$$
where the infimum is over all POVMs $\{E_m\}$, and $p_m={\rm
Tr}(\rho E_m)$, $q_m={\rm Tr}(\sigma E_m)$ are the probability
distributions for $\rho$ and $\sigma$ corresponding to the POVM
$\{E_m\}$. Moreover, the infimum attains the minimum if and only if
the operator
$M=\rho^{[-1/2]}\sqrt{\rho^{1/2}\sigma\rho^{1/2}}\rho^{[-1/2]}$ (may
unbounded) is diagonal.}

Firstly, we need a lemma.

{\bf Lemma 3.2.}  {\it Let $H$ be an infinite dimensional complex
Hilbert space.  Assume that $A\in{\mathcal T}(H)$ and $ \{
T_n\}_{n=0}^\infty\subset{\mathcal B}(H)$. If
SOT-$\lim_{n\rightarrow\infty}T_n=T_0$, then
$\lim_{n\rightarrow\infty}{\rm Tr}(T_nA)={\rm Tr}(T_0A)$. Here SOT
means the strong operator topology. }

{\bf Proof.}  As $T_n$ converges to $T_0$ under the strong operator
topology, there is a constant $d>0$ such that $\sup_n\|T_n\|\leq d$.
For any $\varepsilon
>0$, there exists a finite rank projection $P_\varepsilon$ such that
$\|A-P_\varepsilon AP_\varepsilon\|_{\rm Tr}<\varepsilon/(2d+1)$
because $A\in{\mathcal T}(H)$. On the other hand, $P_\varepsilon$ is
of finite rank, together with SOT-$\lim_{n\rightarrow\infty}
T_n=T_0$, implies that
$$\lim_{n\rightarrow\infty}\|P_\varepsilon(T_n-T_0)P_\varepsilon\|=0.$$
So, for above $\varepsilon>0$, there exists some $N$ such that
$$\|P_\varepsilon(T_n-T_0)P_\varepsilon\|<\frac{1}{(2d+1)\|A\|_{{\rm Tr}}}\varepsilon$$whenever
$n>N$.  Thus we have
$$\begin{array}{rl}|{\rm Tr}((T_n-T_0)A)|
&\leq|{\rm Tr}((T_n-T_0)(A-P_\varepsilon AP_\varepsilon))|+|{\rm Tr}((T_n-T_0)P_\varepsilon AP_\varepsilon)|\\
&\leq\|T_n-T_0\|\|A-P_\varepsilon AP_\varepsilon\|_{{\rm
Tr}}+\|P_\varepsilon(T_n-T_0)P_\varepsilon\|\|A\|_{{\rm
Tr}}\\
&<2d\|A-P_\varepsilon AP_\varepsilon\|_{{\rm
Tr}}+\frac{\varepsilon}{2d+1}\\
&<\frac{2d}{2d+1}\varepsilon
+\frac{\varepsilon}{2d+1}=\varepsilon.\end{array}$$ Therefore,
$\lim_{n\rightarrow\infty}{\rm Tr}(T_nA)={\rm Tr}(T_0A)$.
\hfill$\Box$

{\bf Proof of Theorem 3.1.} Let $\{E_m\}$ be a POVM. Then $E_m\geq
0$ and $\sum_m E_m=I$, here the series converges under the strong
operator topology. By Lemma 2.1, there exists a unitary operator $U$
such that
$\sqrt{\rho^{1/2}\sigma\rho^{1/2}}=U\sqrt{\sigma}\sqrt{\rho}$. Thus,
by the Cauchy-Schwarz inequality and Lemma 3.2,
$$ \begin{array}{rl}
F(\rho,\sigma)= & {\rm Tr}(U\sqrt{\sigma}\sqrt{\rho})=\sum_m{\rm
Tr}(U\sqrt{\sigma}\sqrt{E_m}\sqrt{E_m}\sqrt{\rho}) \\
\leq & \sum_m\sqrt{{\rm Tr}(\rho E_m){\rm Tr}(\sigma E_m)}
=\sum_m\sqrt{p_mq_m}=F(p_m,q_m).\end{array}\eqno(3.3)
$$ Hence we have
$$F(\rho,\sigma)\leq\inf\limits_{\{E_m\}} F(p_m,q_m).$$

Next we show that the equality holds in the above inequality, that
is, Eq.(3.2) holds. By the spectral decomposition, there  is an
orthonormal basis $\{|i\rangle\}_{i=1}^\infty$ of $H$ such that
$\rho=\sum_{i} r_i|i\rangle\langle i|$ with $\sum_i r_i=1$. For any
positive integer $n$, let $H_n$ be the $n$-dimensional subspace
spanned by $|1\rangle, |2\rangle,\ldots , |n\rangle$,  and  $P_n$ be
the projection from $H$ onto $H_n$. Define $\rho _n=\alpha_n^{-1}
P_n\rho P_n$ and $\sigma_n=\beta_n^{-1} P_n\sigma P_n$, where
$\alpha_n={\rm Tr}(P_n\rho P_n)$ and $\beta_n={\rm Tr}(P_n\sigma
P_n)$. Clearly, $\lim_{n\rightarrow\infty} \alpha_n=1$,
$\lim_{n\rightarrow\infty} \beta_n=1$,
SOT-$\lim_{n\rightarrow\infty}
\rho_n=$SOT-$\lim_{n\rightarrow\infty} P_n\rho P_n=\rho$ and
SOT-$\lim_{n\rightarrow\infty}
\sigma_n=$SOT-$\lim_{n\rightarrow\infty} P_n\sigma P_n=\sigma$. By
\cite{ZM}, we see that $\lim_{n\rightarrow\infty} \rho_n=\rho$ and
$\lim_{n\rightarrow\infty} \sigma_n=\sigma$ under the trace norm
topology. It follows that
$\lim_{n\rightarrow\infty}\sqrt{\rho_n^{1/2}\sigma_n\rho_n^{1/2}}=\sqrt{\rho^{1/2}\sigma\rho^{1/2}}$
 under the trace norm
topology, which implies that
$\lim_{n\rightarrow\infty}F(\rho_n,\sigma_n)=F(\rho,\sigma)$. So,
for any $\varepsilon>0$, there exists some $N_1$ such that
$$|F(\rho,\sigma)-\alpha_n\beta_nF(\rho_n,\sigma_n)|<\varepsilon/2 \eqno(3.4)$$
whenever  $n>N_1$.

On the other hand, note that
$\lim_{n\rightarrow\infty}\sqrt{\alpha_n\beta_n}{\rm Tr}(\rho
P_n)=1$ and $\lim_{n\rightarrow\infty}\sqrt{\alpha_n\beta_n}{\rm
Tr}(\sigma P_n)=1$. Thus, for the  above $\varepsilon>0$, there
exists some $N_2$ such that
$$|1-\sqrt{\alpha_n\beta_n}{\rm
Tr}(\rho P_n)|<\varepsilon/2\quad{\rm and}\quad
|1-\sqrt{\alpha_n\beta_n}{\rm Tr}(\sigma P_n)|<\varepsilon/2
\eqno(3.5)$$ whenever  $n>N_2$.

Now, consider $\rho_n$ and $\sigma_n$ for $n\geq \max\{N_1,N_2\}$.
With respect to the space decomposition $H=H_n\oplus H_n^\bot$, we have $\rho_n=\left(\begin{array}{cc} \rho_0 &0\\
0& 0\end{array}\right)$ and $\sigma_n=\left(\begin{array}{cc} \sigma_0 &0\\
0& 0\end{array}\right)$, where $\rho_0,\sigma_0\in{\mathcal
S}(H_n)$. Applying Eq.(3.1) to $\rho_0$ and $\sigma_0$, there exists
POVM $\{E_m'\}\subseteq{\mathcal B}(H_n)$ with
$\sum_{m=1}^nE_m'=I_n$ such that
$$F(\rho_0,\sigma_0)=\sum_{m=1}^n\sqrt{{\rm Tr}(\rho_0 E_m'){\rm Tr}(\sigma_0 E_m')}.$$
Let $E_m=E_m'\oplus 0$ and $E_{n+1}=I-P_n$. It is obvious that
$\sum_{m=1}^{n+1}E_m=I$ and
$$F(\rho_n,\sigma_n)=F(\rho_0,\sigma_0)
=\sum_{m=1}^n\sqrt{{\rm Tr}(\rho_0 E_m'){\rm Tr}(\sigma_0
E_m')}=\sum_{m=1}^{n+1}\sqrt{{\rm Tr}(\rho_n E_m){\rm Tr}(\sigma_n
E_m)}.$$

Now define $F_m=\sqrt{\alpha_n\beta_n}P_nE_mP_n$ for
$m=1,2,\cdots,n+1$ and $F_0=I-\sqrt{\alpha_n\beta_n}P_n$. It is
clear that $\{F_m\}$ is a POVM. Furthermore
$$\begin{array}{rl}\sum_{m=1}^{n+1}\sqrt{{\rm Tr}(\rho F_m){\rm Tr}(\sigma F_m)}
=&\sum_{m=1}^{n+1}\sqrt{{\alpha_n\beta_n}{\rm Tr}(P_n\rho P_n
E_m){\rm
Tr}(P_n\sigma P_nE_m)}\\
=&\sum_{m=1}^{n+1}\sqrt{\alpha_n\beta_n}\sqrt{\alpha_n{\rm
Tr}(\rho_n
E_m)\beta_n{\rm Tr}(\sigma_nE_m)}\\
=&\sum_{m=1}^{n+1}\alpha_n\beta_n\sqrt{{\rm Tr}(\rho_n E_m){\rm
Tr}(\sigma_nE_m)}\\
=&\alpha_n\beta_nF(\rho_n,\sigma_n).\end{array}\eqno(3.6)$$ Hence,
by Eqs.(3.4)-(3.6), we get
$$\begin{array}{rl}
&|F(\rho,\sigma)-\sum_{m=0}^{n+1}\sqrt{{\rm Tr}(\rho F_m){\rm
Tr}(\sigma F_m)}|\\
 \leq&|F(\rho,\sigma)-\sum_{m=1}^{n+1}\sqrt{{\rm
Tr}(\rho F_m){\rm Tr}(\sigma F_m)}|+\sqrt{{\rm Tr}(\rho F_0){\rm
Tr}(\sigma F_0)}\\
=&|F(\rho,\sigma)-\alpha_n\beta_nF(\rho_n,\sigma_n)|+\sqrt{(1-\sqrt{\alpha_n\beta_n}{\rm
Tr}(\rho P_n))(1-\sqrt{\alpha_n\beta_n}{\rm Tr}(\sigma P_n))}\\
<&\varepsilon/2+\varepsilon/2=\varepsilon.\end{array}$$ Thus we have
proved that, for any $\varepsilon>0$, there exists some POVM
$\{F_m\}$ such that
$$F(p_m,q_m)<F(\rho,\sigma)+\varepsilon,$$ where $p_m={\rm
Tr}(\rho F_m)$, $q_m={\rm Tr}(\sigma F_m)$ are the probability
distributions for $\rho$ and $\sigma$ corresponding to the POVM
$\{F_m\}$. So Eq.(3.2) is true.

It is clear that the infimum  of Eq.(3.2) attains the minimum if and
only if there exists a POVM $\{E_m\}$ such that the Cauchy-Schwarz
inequality is satisfied with equality for each term in the sum of
Eq.(3.3), that is,
$$\sqrt{E_m}\sqrt{\rho}=\lambda_m\sqrt{E_m}\sqrt{\sigma} U^\dag \eqno(3.7)$$ for
some set of numbers $\lambda_m \geq 0$. Note that
$\sqrt{\rho^{1/2}\sigma\rho^{1/2}}=U\sqrt{\sigma}\sqrt{\rho}=\sqrt{\rho}\sqrt{\sigma}U^\dag$,
we get that the range of $\sqrt{\rho^{1/2}\sigma\rho^{1/2}}$ is
contained in the range of $\rho^{1/2}$ and hence
$$\sqrt{\sigma}U^\dag=\rho^{[-1/2]}\sqrt{\rho^{1/2}\sigma\rho^{1/2}}.
\eqno(3.8)$$ Substituting Eq.(3.8) into Eq.(3.7), we find that
$$\sqrt{E_m}\sqrt{\rho}=\lambda_m\sqrt{E_m}\rho^{[-1/2]}\sqrt{\rho^{1/2}\sigma\rho^{1/2}} \eqno(3.9)$$
for each $m$. It follows that $
\sqrt{E_m}\sqrt{\rho}\not=0\Rightarrow \lambda _m\not=0$. While, if
$\sqrt{E_m}\sqrt{\rho}=0$, one may take $\lambda_m=0$. Let $H_0={\rm
span}\{ {\rm ran}(\sqrt{E_m}): \sqrt{E_m}\sqrt{\rho}=0\}$, and $P_0$
be the projection onto $H_0$. Then,  Eq.(3.9) implies that Eq.(3.7)
is equivalent to
$$\sqrt{E_m}(I-P_0-\lambda_m M)=0 \eqno(3.10)$$
holds for all $m$, where
$M=\rho^{[-1/2]}\sqrt{\rho^{1/2}\sigma\rho^{1/2}}\rho^{[-1/2]}$ (may
unbounded). Now it is easily seen that the closure of
ran($\sqrt{E_m}$) reduces $M$ to the scalar operator
$\lambda_m^{-1}$ if $\sqrt{E_m}\sqrt{\rho}\not=0$, and $\ker M=H_0$.
Thus $0,\lambda_m^{-1}\in\sigma_p(M)$, the point spectrum (i.e.,
eigenvalues) of $M$. Since $\sum_m E_m=I$, we see that $\sum_m {\rm
ran} (\sqrt{E_m})=H$ and the spectrum of $M$,
$\sigma(M)\subseteq{\rm cl} \ \{0,\lambda_m^{-1}\}={\rm cl}\
\sigma_p(M)$. So $M$ must be diagonal. Conversely, if $M$ is
diagonal, say $M=\sum_m \gamma_m |m\rangle\langle m|$ with
$\{|m\rangle\}$ an orthonormal basis of $H$. Let
$\lambda_m=\gamma_m^{-1}$ if $\gamma_m\not=0$; $\lambda_m=0$ if
$\gamma_m=0$. Then the POVM $\{E_m=|m\rangle\langle m|\}$ satisfies
Eq.(3.10) and thus Eq.(3.9). Hence $F(\rho,\sigma)=\sum_m\sqrt{{\rm
Tr}(\rho E_m){\rm Tr}(\sigma E_m)} =F(p_m,q_m)$. This completes the
proof.\hfill$\Box$

{\bf Remark 3.3.} There do exist some $\rho$ and $\sigma$ such that
there is no POVM $\{E_m\}$ satisfying
$F(\rho,\sigma)=\sum_m\sqrt{{\rm Tr}(\rho E_m){\rm Tr}(\sigma E_m)}
$. For example, let $H=L_2([0,1])$ and $M_t$ the operator defined by
$(M_tf)(t)=tf(t)$ for any $f\in H$. Then $M_t$ is positive and is
not diagonal because $\sigma(M_t)=[0,1]$ and the point spectrum
$\sigma_p(M_t)=\emptyset$. Let $\rho\in{\mathcal S}(H)$ be injective
as an operator. Then $d={\rm Tr}(M_t^2\rho)\not=0$. Let
$M=d^{-1}M_t$ and $\sigma =M\rho M$. As ${\rm Tr}(M^2\rho)=1$,
$\sigma$ is a state. Now it is clear that
$M=\rho^{-1/2}\sqrt{\rho^{1/2}\sigma\rho^{1/2}}\rho^{-1/2}$, which
is not diagonal. Thus by Theorem 3.1, the infimum in Eq.(3.2) does
not attain the minimum.

For two states $\rho$ and $\sigma$, recall that the trace distance
of them is defined by
$D(\rho,\sigma)=\frac{1}{2}\|\rho-\sigma\|_{\rm Tr}$. By use of
Uhlmann's theorem and Eq.(3.1), it holds for finite dimensional case
that
$$1-F(\rho,\sigma)\leq D(\rho,\sigma)\leq
\sqrt{1-F(\rho,\sigma)^2}. \eqno(3.11)$$ This reveals that the trace
distance and the fidelity are qualitatively equivalent measures of
closeness for quantum states. Now Theorem 3.1 allows us to establish
the same relationship between fidelity measure and trace distance
measure for states of infinite dimensional systems.

{\bf Theorem 3.4.} {\it Let  $H$ be an infinite dimensional
separable complex Hilbert space. Then for  any states
$\rho,\sigma\in{\mathcal S}(H)$, the inequalities in Eq.(3.11)
hold.}

{\bf Proof.} Firstly, it is obvious that if both
$\rho=|a\rangle\langle a|$ and $\sigma=|b\rangle\langle b|$ are pure
states, then $D(\rho,\sigma)=D(|a\rangle,
|b\rangle)=\sqrt{1-F(|a\rangle,|b\rangle)^2}=\sqrt{1-F(\rho,\sigma)^2}.
$ (Ref. \cite[pp. 415]{NC} for a proof that is valid for both finite
and infinite dimensional cases.)

Let $\rho$ and $\sigma$ be any two states, and let $|\psi\rangle$
and $|\phi\rangle$  be purifications chosen such that
$F(\rho,\sigma)=|\langle\psi|\phi\rangle |$ by Theorem 2.4. Since
the trace distance is non-increasing under the partial trace, we see
that
$$ D(\rho,\sigma)\leq D(|\psi\rangle,
|\phi\rangle)=\sqrt{1-F(|\psi\rangle,|\phi\rangle)^2}=\sqrt{1-F(\rho,\sigma)^2}.$$
This establishes the inequality
$$D(\rho,\sigma)\leq \sqrt{1-F(\rho,\sigma)^2}. \eqno(3.12)$$

To see the other inequality of Eq.(3.11)  is true, Theorem 3.1 is
needed.

For any given $\varepsilon>0$, by Theorem 3.1, we may take a POVM
$\{E_m\}$ such that
$$ F(\rho,\sigma)\leq F(p_m,q_m)=\sum_m\sqrt{p_mq_m}< F(\rho,\sigma)+\varepsilon, \eqno(3.13)$$
where $p_m={\rm Tr}(\rho E_m)$ and $q_m={\rm Tr}(\sigma E_m)$ are
the probabilities for obtaining outcome $m$ for the states $\rho$
and $\sigma$, respectively. Observe that, for both finite and
infinite dimensional cases, we have
$$ D(\rho,\sigma)=\max\limits_{\{E_m\}} D(p_m,q_m),\eqno(3.14)$$
where $D(p_m,q_m)=\frac{1}{2}\sum_m |p_m-q_m|$ and the maximun is
over all POVM $\{E_m\}$. It follows from Eq.(3.14) and
$$ \sum_m(\sqrt{p_m}-\sqrt{q_m})^2=\sum_mp_m+\sum_mq_m-2F(p_m,q_m)
=2(1-F(p_m,q_m)),$$ that
$$\begin{array}{rl} 2(1-F(\rho,\sigma))-2\varepsilon <&
2(1-F(p_m,q_m))=\sum_m(\sqrt{p_m}-\sqrt{q_m})^2 \\ \leq & \sum_m
|\sqrt{p_m}-\sqrt{q_m}|(\sqrt{p_m}+\sqrt{q_m})=\sum_m |p_m-q_m|
\\=&2D(p_m,q_m)\leq 2D(\rho,\sigma). \end{array}
$$
Thus we have proved that
$$(1-F(\rho,\sigma))-\varepsilon <D(\rho,\sigma)$$
holds for any $\varepsilon>0$.  This forces that
$$1-F(\rho,\sigma)\leq D(\rho,\sigma),$$ which, combining the
inequality (3.12), completes the proof of the theorem.\hfill$\Box$

\section{Fidelities connected to channels}

For finite dimensional case, ensemble average fidelity and
entanglement fidelity are two kinds of important fidelities
connected to a quantum channel.  In this section we give the
definitions of ensemble average fidelity and entanglement fidelity
connected to a quantum channel for an infinite dimensional system,
and discuss their relationship.

 Let $H$ be an infinite dimensional separable complex
Hilbert space. Recall that a quantum channel ${\mathcal E}:
{\mathcal T}(H)\rightarrow{\mathcal T}(H)$ is a trace preserving
completely positive linear map. Like the finite dimensional case,
for such quantum channel ${\mathcal E}$ and a given ensemble
$\{p_j,\rho_j\}_{j=1}^\infty$, one can define ensemble average
fidelity by
$$\overline{F}=\sum_jp_jF(\rho_j,{\mathcal E}(\rho_j))^2.\eqno(4.1)$$
Similarly, for a state $\rho$, one can define the entanglement
fidelity by
$$\begin{array}{rl}F(\rho,{\mathcal E})=&F(|\psi\rangle,({\mathcal E}\otimes I)(|\psi\rangle\langle\psi|))^2\\
=&\langle\psi|({\mathcal E}\otimes
I)(|\psi\rangle\langle\psi|)|\psi\rangle,\end{array}\eqno(4.2)$$
where $|\psi\rangle\in H\otimes H$ is a purification of $\rho$. Note
that the definition $F(\rho,{\mathcal E})$ does not depend on the
choices of purifications. To see this, let
$|\psi\rangle=\sum_j\sqrt{p_j}|j\rangle|\mu_j\rangle$ be any
purification, where $\{j\}$ is an orthonormal basis  and $\{\mu_j\}$
is an orthonormal set of $H$. By \cite{H}, there exists a sequence
of operators  $\{E_i\}\subseteq {\mathcal B}(H)$ with
$\sum_iE_i^\dag E_i=I$ such that $${\mathcal
E}(\sigma)=\sum_iE_i\sigma E_i^\dag \quad{\rm for\ \ all}\quad
\sigma\in{\mathcal S}(H).$$ Thus
$$\begin{array}{rl}F(\rho,{\mathcal E})=&\sum_i\langle\psi|(E_i\otimes I)(|\psi\rangle\langle\psi|)(E_i^\dag\otimes I)|\psi\rangle\\
=&\sum_i\langle\psi|\sum_{j,k}\sqrt{p_jp_k}(E_i\otimes
I)(|j\rangle|\mu_j\rangle\langle k|\langle\mu_k|)(E_i^\dag\otimes
I) |\psi\rangle\\
=&\sum_i\sum_{j,k}p_jp_k\langle j|E_i|j\rangle\langle
k|E_i^\dag |k\rangle\\
=&\sum_i|{\rm Tr}(E_i\rho)|^2,\end{array}\eqno(4.3)$$ which is
dependent only to $\rho$ and $\mathcal E$.

In the sequel we will give some properties of entanglement fidelity
for infinite dimensional systems.

Firstly note that,  by monotonicity of the fidelity Eq.(2.7), it is
easily checked that
$$F(\rho,{\mathcal E})\leq[F(\rho,{\mathcal E}(\rho))]^2.\eqno(4.4)$$

{\bf Proposition 4.1.} {\it Let  $H$ be an infinite dimensional
separable complex Hilbert space. Assume that ${\mathcal E}:
{\mathcal T}(H)\rightarrow{\mathcal T}(H)$ is a quantum channel and
$\rho\in{\mathcal S}(H)$. Then the entanglement fidelity
$F(\rho,{\mathcal E})$ is a convex function of $\rho$.}

{\bf Proof.} Take any states $\rho_1,\rho_2\in{\mathcal S}(H)$.
Define a real function $f:{\mathbb R}\rightarrow{\mathbb R}$ by
$$f(x)\equiv F(x\rho_1+(1-x)\rho_2,{\mathcal
E}),\quad \forall x\in{\mathbb R}.$$ By using of Eq.(4.3) and
elementary calculus, one sees that the second  derivative of $f$ is
$$f''(x)=\sum_i|{\rm Tr}((\rho_1-\rho_2)E_i)|^2.$$ Hence $f''(x)\geq
0$, which implies that $F(\rho,{\mathcal E})$ is convex, as desired.
\hfill$\Box$

{\bf Proposition 4.2.} {\it Let  $H$ be an infinite dimensional
separable complex Hilbert space. Assume that ${\mathcal E}:
{\mathcal T}(H)\rightarrow{\mathcal T}(H)$ is a quantum channel.
Then for any given ensemble $\{p_j,\rho_j\}$, we have
$F(\sum_jp_j\rho_j,{\mathcal E})\leq \overline{F}$.}

{\bf Proof.} For any $k\in{\mathbb N}$, let
$\lambda_k=\sum_{j=1}^kp_j$. Then, by Proposition 4.1, we have
$$\begin{array}{rl}F(\sum_jp_j\rho_j,{\mathcal E})=&
F(\lambda_k(\sum_{j=1}^k\frac{p_j}{\lambda_k}\rho_j)+(1-\lambda_k)(\sum_{j=k+1}^\infty\frac{p_j}{1-\lambda_k}\rho_j),{\mathcal
E})\\
\leq& \lambda_k F(\sum_{j=1}^k\frac{p_j}{\lambda_k}\rho_j,{\mathcal
E})+(1-\lambda_k)F(\sum_{j=k+1}^\infty\frac{p_j}{1-\lambda_k}\rho_j,{\mathcal
E})\\
\leq& \lambda_k \sum_{j=1}^k\frac{p_j}{\lambda_k}F(\rho_j,{\mathcal
E})+(1-\lambda_k)F(\sum_{j=k+1}^\infty\frac{p_j}{1-\lambda_k}\rho_j,{\mathcal
E})\\
=&\sum_{j=1}^k{p_j}F(\rho_j,{\mathcal
E})+(1-\lambda_k)F(\sum_{j=k+1}^\infty\frac{p_j}{1-\lambda_k}\rho_j,{\mathcal
E}). \end{array}\eqno(4.5)$$ Note that $0\leq F(\rho,{\mathcal
E})\leq 1$ and $\lim_{k\rightarrow\infty}\lambda_k=\sum_{j=1}^\infty
p_j=1$. So
$$\lim_{k\rightarrow\infty}(1-\lambda_k)F(\sum_{j=k+1}^\infty\frac{p_j}{1-\lambda_k}\rho_j,{\mathcal
E})=0.$$ Thus, for any $\varepsilon>0$, there exists some $N$ such
that
$$(1-\lambda_k)F(\sum_{j=k+1}^\infty\frac{p_j}{1-\lambda_k}\rho_j,{\mathcal
E})<\varepsilon\eqno(4.6)$$ whenever  $k>N$. It follows from
Eq.(4.5) that
$$F(\sum_jp_j\rho_j,{\mathcal E})<\sum_{j=1}^\infty{p_j}F(\rho_j,{\mathcal
E})+\varepsilon.$$ By the arbitrariness of $\varepsilon$ and
Eq.(4.4), we obtain that
$$\begin{array}{rl}F(\sum_jp_j\rho_j,{\mathcal E})\leq&\sum_{j=1}^\infty{p_j}F(\rho_j,{\mathcal
E})\\
\leq&\sum_{j=1}^\infty{p_j}F(\rho_j,{\mathcal
E}(\rho_j))^2=\overline{F},\end{array}$$ Completing the proof.
\hfill$\Box$

\section{Conclusion}

In this paper we prove the infinite dimensional version of the
Uhlmann's theorem by an elementary approach, which states that the
fidelity of states $\rho$ and $\sigma$ is   larger than or
 equal to the absolute value of the inner product of any purifications $|\psi\rangle$ and $|\phi\rangle$ of $\rho$ and
 $\sigma$,
 i.e., $F(\rho,\sigma)\geq |\langle\psi|\phi\rangle|$; moreover, there exist some purifications such that the equality
 holds. This allows us to generalize a large part of the results concerning
 fidelity for finite dimensional systems to that for infinite
 dimensional systems. We also
discuss the relationship between quantum fidelity and classical
fidelity and show that $F(\rho,\sigma)=\inf_{\{E_m\}}F(p_m,q_m)$.
Not like to that of finite dimensional case, the infimum can not
attain the minimum in general. We give a necessary and sufficient
condition for the infimum attains the minimum. Using this result, we
find that the fidelity and the trace distance are equivalent in
describing the closeness of states. The concepts of
 ensemble average  fidelity and  entanglement fidelity for a channel are
 generalized to infinite dimensional case. The relationship of such
 two fidelities is discussed.

\end{document}